\documentclass[12pt,preprint]{aastex}
\usepackage{natbib}

\newcommand{\mc}{\multicolumn}
\newcommand{\expnt}[2]{\ensuremath{#1 \times 10^{#2}}}   
\newcommand{\gsim}{\stackrel{>}{{\sim}}}
\newcommand{\lsim}{\stackrel{<}{\sim}}

\newcommand{\sgr}{SGR~1806$-$20}
\newcommand{\snr}{G10.0$-$0.3}
\newcommand{\chandra}{\textit{Chandra}}
\newcommand{\cxo}{\textit{CXO}}

\shorttitle{Precise \chandra\ Localization of \sgr}
\shortauthors{Kaplan et al.}

\begin{document}

\title{Precise \chandra\ Localization of the Soft Gamma-ray Repeater \sgr}
\author{D.~L.~Kaplan, D.~W.~Fox, S.~R.~Kulkarni}
\affil{Department of Astronomy, 105-24 California Institute of
Technology, Pasadena, CA 91125, USA}
\email{dlk@astro.caltech.edu, derekfox@astro.caltech.edu,
srk@astro.caltech.edu}
\author{E.~V.~Gotthelf}
\affil{Columbia Astrophysics Laboratory, Columbia University, 550 West
120th Street, New York, NY 10027, USA}
\email{evg@astro.columbia.edu}
\author{G.~Vasisht}
\affil{Jet Propulsion Laboratory, California Institute of Technology,
4800 Oak Grove Drive, Pasadena, CA 91109, USA}
\email{gv@huey.jpl.nasa.gov}

\and
\author{D. A. Frail}
\affil{National Radio Astronomy Observatory, Socorro, NM 87801, USA}
\email{dfrail@nrao.edu}

\keywords{pulsars: individual (\sgr) --- stars: neutron
--- X-rays: stars --- astrometry}

\slugcomment{Accepted by ApJ}

\begin{abstract}
We present observations of the Soft $\gamma$-ray Repeater \sgr\ taken with
the \textit{Chandra X-ray Observatory}.  We identify the X-ray
counterpart of \sgr\ based on detection of 7.5-s pulsations.  Using three unrelated X-ray
sources (and USNO stars) as position references, we are able to
determine that the SGR is at $\alpha_{2000}=18^{\rm h}08^{\rm m}39\fs32$
and $\delta_{2000}  =  -20\degr 24\arcmin 39\farcs5$, with rms
uncertainty of $0\farcs3$ in each coordinate.  We find
that \sgr\ is located within the 1-$\sigma$ error region determined by
Interplanetary Network data and is 14$\pm$0.5~arcsec distant from the
non-thermal core of SNR \snr, excluding \sgr\ as the origin of
the core.   We see evidence for a significant
deviation of the spin-down of \sgr\ from its long-term trend,
demonstrating  erratic spin-down behavior
in this source similar to that seen in other SGRs.  Finally, we show
that there is a broad X-ray halo surrounding \sgr\ out to radii $\sim
1\arcmin$ due to scattering in the interstellar medium.
\end{abstract}

\section{Introduction}
The soft $\gamma$-ray repeaters  (SGRs; see \citealt{h99} for a recent
observational review) are enigmatic sources that were discovered
through their repeated bursts of hard X-rays/soft
$\gamma$-rays.  With the improvement in imaging afforded by
the current generation of X-ray satellites SGRs were found to emit  softer
quiescent X-ray emission  as well.   They are 
generally thought to be young ($<10^{4}$~yr) neutron stars with
extremely strong magnetic fields, or magnetars
\citep[e.g.][]{dt92,td93}.  This belief has been motivated by their
associations with young supernova remnants (SNRs;
\citealt{ekl+80,kf93,vkfg94}) and sites of massive star formation
\citep{fmc+99,vhl+00}, the energetics of their bursts
\citep{p92,td95}, and the detection of X-ray pulsations with long
(5--10~s) periods and large ($\sim 10^{-11}\mbox{ s s}^{-1}$)
spin-down rates \citep[e.g.][]{ksh+99}.

The proposed identification of a SNR, the plerion \snr, with \sgr\
by \citet{kf93} played a key role in establishing the 
connection of SGRs with SNRs and thence young stars. This identification
renewed confidence in the association of N49 with the 5~March~1979
event (SGR~0526$-$66) and motivated SNR associations for two of the remaining
 SGRs, SGR~1900+14 \citep{vkfg94} and SGR 1627$-$41
\citep{hkw+99}.   The association of \sgr\ with \snr\
was strengthened by the discoveries of a non-thermal core in
\snr\ (\citealt{kfk+94}; \citealt*{vfk95}) that changed on month to
year time-scales 
and of a nearby luminous blue variable star
\citep[LBV;][]{kmn+95,vkkmn95}.  It was proposed that all of these
sources were related, with the LBV as either a current or past binary
companion to \sgr\ \citep{vkkmn95} and the variable radio source 
the  result of episodic mass loss 
from \sgr\ \citep*{fvk97}.

However, a recent moderate-precision IPN position  \citep{hkc+99} placed \sgr\
outside the non-thermal core suggested by \citet{kfk+94} to be
the seat of the SGR. Furthermore, the association of SGR~1900+14
with SNR~G42.8+0.6  was questioned \citep{lx00}.  In summary, the
entire issue of whether SGRs are associated with SNRs is now open to
debate (see review by \citealt{gsgv01}).

The purpose of this paper is to provide an independent localization of
the quiescent X-ray counterpart of \sgr.  SNR/SGR associations are
useful for establishing independent ages for SGRs; the non-thermal
core to \snr\ provides key energetics constraints; and the LBV star
hints at a binary origin and evolutionary scenario for \sgr.
Determining whether  or not these associations are true is therefore
key to assessing the nature of \sgr.

\section{Observations \& Analysis}
\label{sec:obs}
We observed \sgr\ with the {\em Chandra X-ray Observatory} on two
occasions, beginning on 2000~July~24.7~UT and 2000~August~15.7~UT,
with durations of 4.9~ks and 31~ks respectively (see
Table~\ref{tab_sum}).  Both observations were taken with the aim point
on the backside-illuminated ACIS S-3 detector.  The first observation
was taken in the standard full-frame CCD mode with 3.24-s time
resolution, while the second observation was acquired
in a 1/4-subarray mode that provides increased time resolution of
0.81~s, but with a reduced field-of-view (1/4 of the full-frame area).
All the data 
suffer from the effects of photon pile-up: $\approx 50$\% of the data
from the first observation and $\lsim 10$\% of the data from the
second are corrupted.  Pile-up occurs when ``two or more photons are
detected as a single
event''\footnote{\url{http://asc.harvard.edu/udocs/docs/POG/MPOG/node11.html\#SECTION046160000000000000000}}.
This results in 3 main effects: (1) the detected event has an energy that
is the sum of all of the incident photon energies; (2) the count rate
is diminished; and (3) the grades of the incoming photons will be
altered, so that some may be cataloged as ``cosmic rays'' or other
undesireable events and will therefore be rejected.  The consequences
of these effects on an
observation of a bright source, in addition to modification of the
spectrum, are that the spatial profile is altered, as the diminishment
of the count-rate is worst for the pixels at the center of the psf,
 and that some regions with especially
high rates of incoming photons may appear devoid of all events, as
the total amplitude of the detected events exceeds the onboard energy
threshold and is rejected (similar to grade migration).

Because of pile-up, especially effects (1) and (2) above, we cannot
perform  spectroscopy 
with great accuracy and we defer such analysis until tools for
correcting pile-up become  publicly available.  As noted, pile-up will also
affect the spatial profile of the source (it depresses the center of
the psf relative to the wings) and the timing (two photons arriving
together will be counted as one) but the
analysis presented here should not be affected significantly, as we
are not interested in the detailed shape of the psf within $1\arcsec$
or in the exact lightcurve --- overall, the position and period will
be preserved.

\begin{figure*}
\plotone{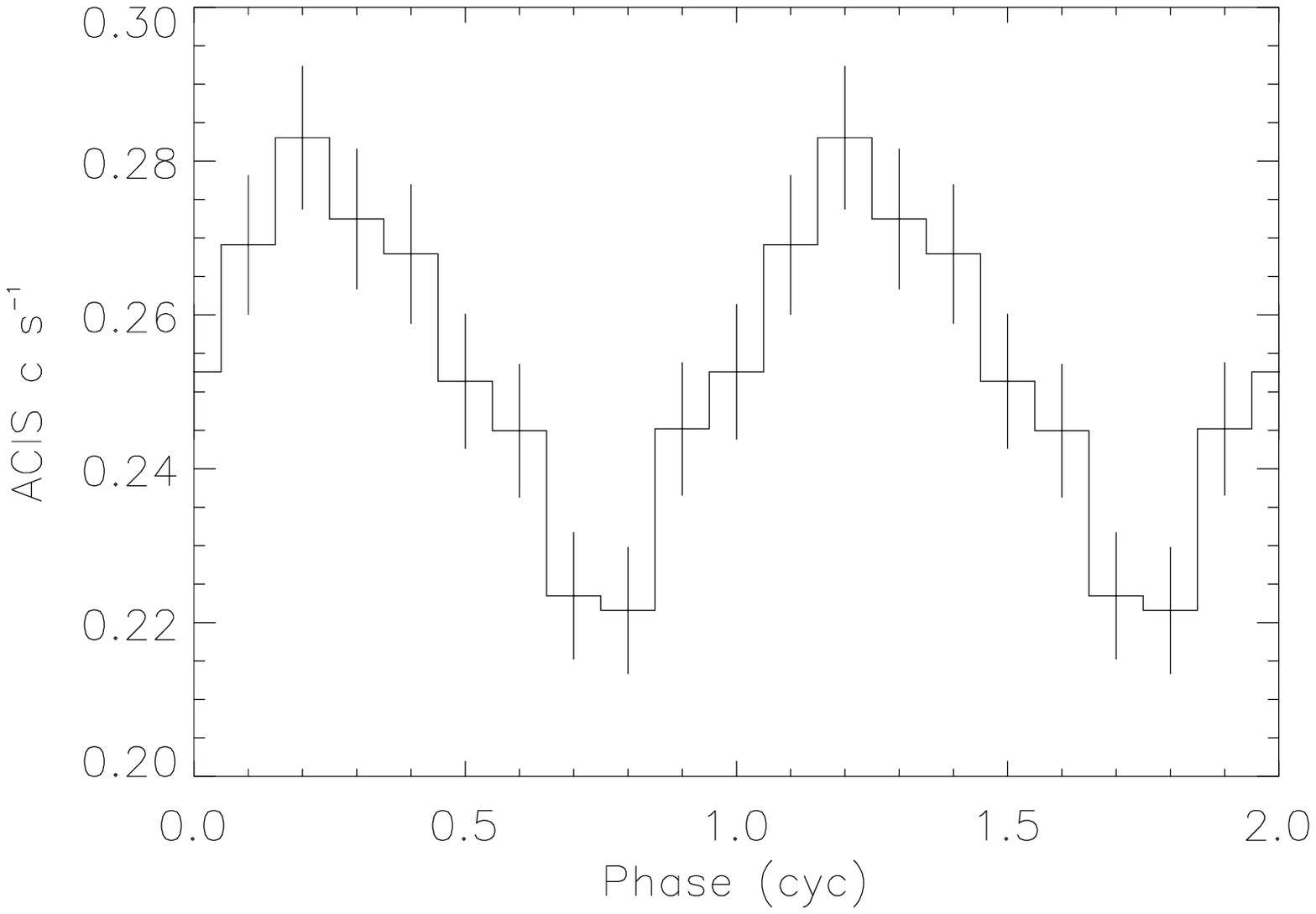}
\caption{Pulse profile for \sgr\ from the 31-ks observation, folded at
the best period of 7.4925~s.  The profile is shown over two periods
for clarity.  No corrections for pile-up have been made.
\label{fig:pulse}}
\end{figure*}

 We examined the two images and find a single bright source near the
expected location of the SGR. This source is consistent with an unresolved
point source (Gaussian $\sigma=0\farcs33$; \citealt{mrf+01}). In order
to determine if this source is indeed the SGR, we searched for the
expected 7-s pulsations \citep{kds+98}.  We first barycentered
the data using the {\tt axBary}  software.  The 5-ks observation had
too few photons to detect anything, and it was taken with 3.24~s time
resolution.  The 31-ks data had sufficient photons and was taken in
the 1/4~subarray mode, which gives 0.81~s sampling.  We therefore
added a random number $\sim {\cal U}(0,0.81)$ to the time-of-arrival
values to eliminate any  effects of sampling.  We then  performed a
$Z_{1}^{2}$ test \citep*{drs89} on the 31-ks data, and find a very
significant periodicity at $P\approx 7.5$~s. To refine this
measurement, we performed a 
phase connection of the \chandra\ data in the manner of
\citet{fkkf01}.  We connected four segments of $\approx 8000$-s, each
of which had been binned to 1-s resolution (appropriate for the
0.8-s sampling).  We find, referenced to MJD~51772.0 (TDB), a phase of
$0.12(5)$~cycles, a period $P=7.4925(2)$~s, and a period derivative
$|\dot{P}| < \expnt{4}{-8}\mbox{ s s}^{-1}$ at 90\% confidence.  The
pulsations have a sinusoidal profile and a rms pulsed fraction of  $\approx 
7.4$\% --- a folded pulse profile is shown in Figure~\ref{fig:pulse}.

Although the count-rate and spectrum from our observations are
uncertain due to pile-up, we can roughly estimate the source 
flux through use of previously published spectra for \sgr\
\citep{mcft00}.  With the count-rate from the 31-ks observation (the
least corrupted), we see 0.5--10~keV fluxes of $\approx
\expnt{7}{-12}\mbox{ erg 
s}^{-1}\mbox{ cm}^{-2}$ (absorbed) and $\approx \expnt{2}{-11}\mbox{ erg
s}^{-1}\mbox{ cm}^{-2}$ (unabsorbed).  These values are entirely
compatible with the 2--10~keV fluxes found by \citet{mcft00}.  The
observed count-rates are constant throughout the observations.

Given the 0.81-s sampling and the presence of pile-up, it is not
surprising that we do not see any significant bursts of the type
typically seen during observations of SGRs \citep[e.g.][]{fkkf01}.
We note that it is likely any bursts during 
this observation were probably not recorded  due to pile-up and to
on-board/post-processing rejection of anomalous events.

\subsection{Localization}
To accurately localize  \sgr, we measured its position in both
\chandra\ datasets.  The measured positions from the two
observations are consistent to $0\farcs04$, suggesting that the
stochastic position errors are minimal.  There may however be
systematic errors on the order of $1\arcsec$ due to overall
uncertainties in the \cxo\ aspect solutions.

To correct for such errors we searched the \chandra\ images
for background sources to use as position references.  We found seven
weak sources in 
the 5-ks image. 
None had a match in the SIMBAD database, but two sources (detected at
2.5-$\sigma$ confidence)  were coincident with stars.    In the 31-ks
image we found 11 sources, two of 
which had matches in the USNO-A2.0 catalog: 
one detected at 4-$\sigma$ confidence that was also in the 5-ks image,
the other detected at 2.5-$\sigma$ confidence.  We summarize the X-ray
detected USNO stars in 
Table~\ref{tab:usno}.  There are $\approx 4.3\mbox{ USNO stars arcmin}^{-2}$
in this region, giving chance coincidence rates of $\sim 0.1$\%
between an individual X-ray source and a USNO star, supporting our
identifications.  We are pursuing 
photometric and spectroscopic observations of these stars that should
allow us to classify them and verify these identifications
\citep[e.g.][]{vdbv01}. 

Using the USNO stars to derive offsets for the astrometry, we find
 corrections of $\Delta \alpha=0\farcs4$ and $\Delta
 \delta=0\farcs5$, consistent between both \chandra\ images and
 comparable in magnitude to the expected aspect uncertainties\footnote{\url{http://asc.harvard.edu/mta/ASPECT/celmon/}}. We find a corrected position
for \sgr\ of $\alpha_{2000} =  18^{\rm h}08^{\rm m}39\fs32$ and 
$\delta_{2000} = -20\degr 24\arcmin 39\farcs5$.
This position has rms uncertainties of $0\farcs3$ in each coordinate
(from centroiding the X-ray sources and intrinsic USNO uncertainties
of $0\farcs2$; \citealt{d99}),
but should be free from systematic uncertainties.
 
\begin{figure*}
\plotone{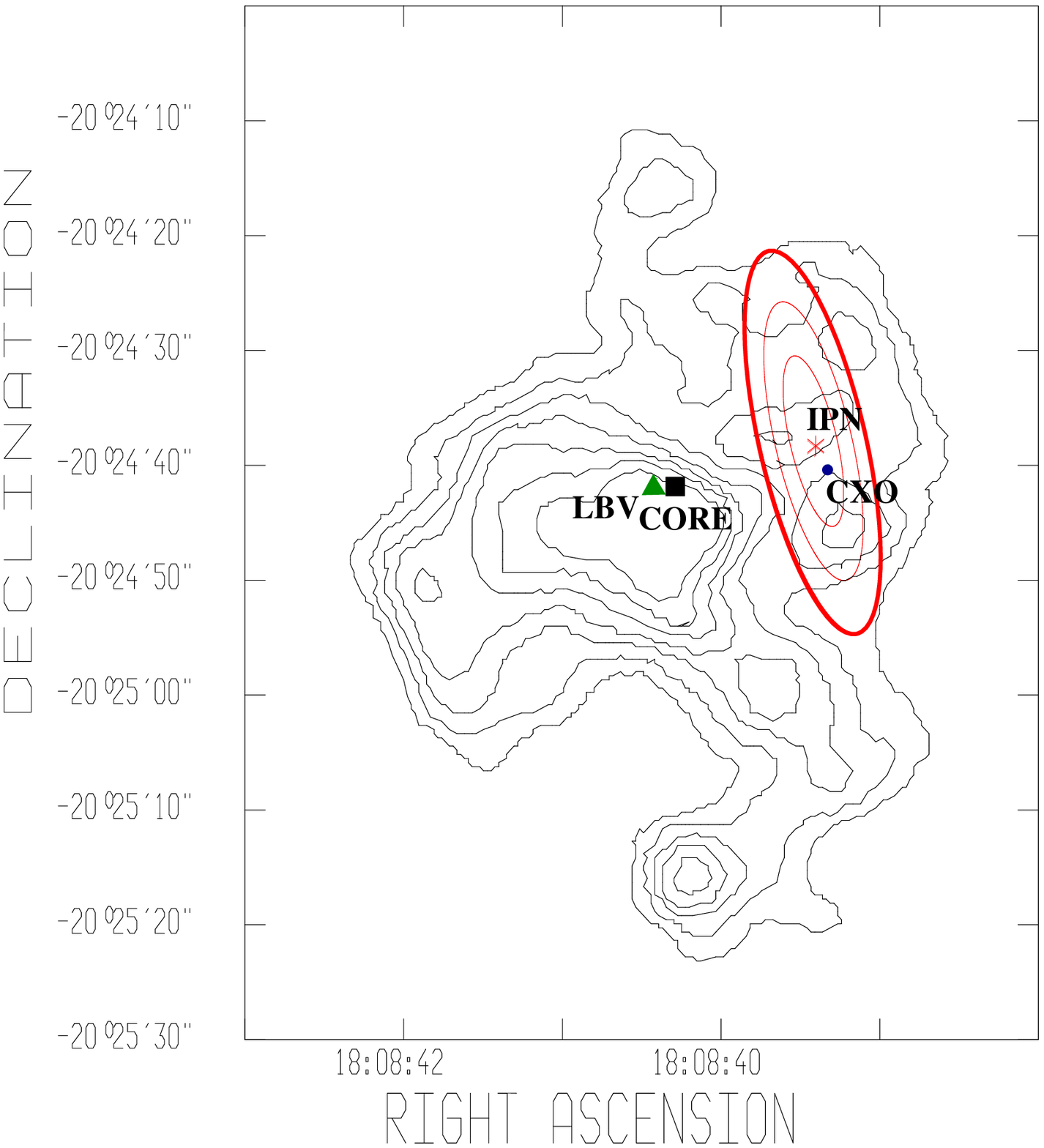}
\caption{3.6-cm VLA image of central portion of \snr\ \citep{fvk97} with position
of radio core at square marked ``CORE''.  Superimposed are 1-, 2-, and
3-$\sigma$ annuli around the best-fit IPN position (asterisk labeled ``IPN'')
from \citet{hkc+99}, the corrected \chandra\ position (circle labeled
``CXO''), and the position of the LBV star (triangle marked ``LBV'';
\citealt{kmn+95}).  Note that the position of the core indicated in
\citet{hkc+99} is slightly incorrect (K.~Hurley 2001, personal
communication).  Adapted from \citet{hkc+99}. 
\label{fig:radio}}
\end{figure*}

This position agrees very well (within the
1-$\sigma$ error ellipse) with the position determined from IPN
measurements \citep{hkc+99}, and is $14\pm 0.5\arcsec$ from the
position of the non-thermal nebula from \citet{fvk97}; see
Figure~\ref{fig:radio}.

\subsection{Extended Emission}
\label{sec:halo}
X-rays are scattered by dust grains in the interstellar medium and one
expects to see halos of size 0.1--100~arcmin towards objects with
hydrogen column density of $10^{21}\mbox{ cm}^{-2}$ or greater
\citep[i.e.,][]{o65,ps95}. This issue was first discussed 
by \citet{o65} and the \textit{Einstein Observatory} found the first such 
halos \citep{r83,c83}. The most comprehensive work to date has been carried out
by Predehl and associates using the \textit{ROSAT Observatory} \citep{ps95}.

\begin{figure*}
\plotone{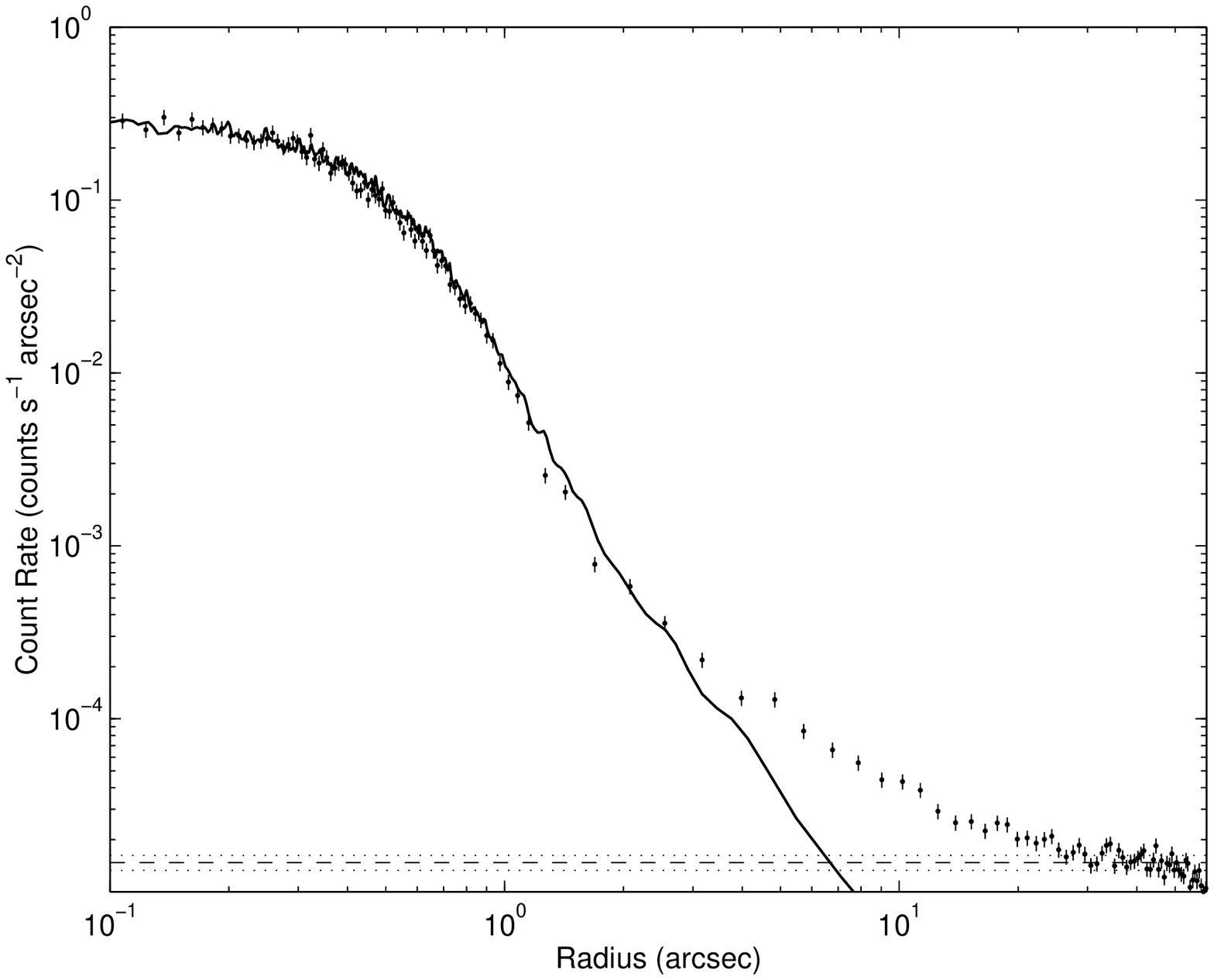}
\caption{Radial profile of \sgr\ from the 31-ks
observation (points), 
along with that of a \texttt{MARX} model of the \chandra\ psf (solid
line) scaled to the same normalization for $r \lsim 1\arcsec$.  The
mean background level (corrected for vignetting) is shown by the
dashed line, with $\pm 1\sigma$ levels indicated by the dotted lines.
\label{fig:halo}}
\end{figure*}

The recent interest of scattering halos around SGRs has been
motivated by the discovery of such a halo around SGR~1900+14 \citep{ktw+01}.
As with SGR~1900+14, one expects an X-ray halo around
\sgr\ from scattering off dust.  We therefore examined the radial
profile of \sgr\  from the 31-ks observation for better
signal-to-noise.  
In Figure~\ref{fig:halo}, we show the
radial profile of \sgr\ from this observation
along with the profile from a
\texttt{MARX}\footnote{\url{http://space.mit.edu/ASC/MARX/}}
simulation (using the 
spectrum for \sgr\ from \citealt{mcft00}) scaled to the same
normalization for the inner portion \citep[cf.][]{ktw+01}; such
simulations should be  accurate out to radii of a few
arcminutes\footnote{\url{http://asc.harvard.edu/cal/Hrma/hrma/psf/psfwings/psfwings.html}}.
The background was measured from the ACIS S-3 chip far ($>3\arcmin$)
from the \sgr, and includes both instrumental and
unresolved-source contributions.  We  corrected the background estimate to
$r \lsim 1\arcmin$ with  a \texttt{MARX} simulation
of the ACIS S-3 response to account for vignetting etc.  We find the
variation in radial response to be small for $r \lsim 1\arcmin$, and
therefore show a constant background in Figure~\ref{fig:halo}.
One sees very good agreement between the data and the model at small
radii, showing that pile-up has  not substantially  corrupted the
radial profile of this observation (the model did not incorporate
pile-up, as \texttt{MARX} does not correctly simulate the
back-illuminated CCDs).  But there is a clear
deviation from the model psf at large radii ($\gsim 3\arcsec$) that we
believe to be indicative of a dust halo contributing $\sim 0.5$\% to
the total X-ray flux.  To verify this interpretation we examined the
radial profile as a function of the phase of the \sgr.  The 
profile for the portion of the 
period when \sgr\ is ``ON'' has of course more counts at small radii
than the profile when \sgr\ is ``OFF.''  However, the outer portion
($r \gsim 3\arcsec$) that we identify as the halo is identical between
 ``ON'' and ``OFF'', indicating that the pulsations of \sgr\ are 
smeared out in the halo.  This suggests
that the extended emission is not an instrumental property, as
scattering wings or something similar would scale with the immediate
flux of the source instead of averaging over time as observed here.
We note, though, that while the data  do show a slight softening
towards higher radii (as expected from the $E^{-2}$
dependence of the scattering cross-section; e.g.\ \citealt{pbpt00}),
we cannot determine in detail how the halo profile changes
with energy.  This could be due to the relatively hard spectrum of
\sgr, so that the cross section does not vary much over the energy
range for which there are significant counts (2--6~keV).  It could
also be an effect of poor statistics, where a longer observation would
show an energy dependent halo.  Regardless, we must caution the reader
as to this interpretation of the extended emission.

\section{Discussion}
The X-ray position of \sgr\ is now without  doubt incompatible with
the positions of both the non-thermal core of \snr\ and the LBV star,
although those two positions still agree to $1\arcsec$.  We therefore
follow \citet{hkc+99} and \citet{gsgv01} in suggesting that the LBV,
not \sgr, powers the core of \snr.  As outlined in \citet{hkc+99} the
LBV can easily explain the energetics and changing morphology of the
non-thermal core.  \sgr\ could still be associated with \snr\ as a
whole, although this is unlikely given the proximities of the core and
the LBV to the center of \snr.

However, we still have a remarkable coincidence:
within a circle $\approx 10\arcsec$ in radius ($0.7$~pc at 14.5~kpc
distance; \citealt{cwd+97}), we have an SGR (\sgr), an LBV, 
and a massive star cluster \citep{fmc+99}.  All three of these objects
are exceedingly rare.  In addition, there is a SNR\footnote{Although
\citet{gsgv01} have argued that \snr\ is not a SNR.} at the same
position.  While it is hard to draw quantitative conclusions from this
coincidence since none of the objects involved have well-defined
populations and this region
of the sky has been studied in great depth, it is still noteworthy.
Individual distances to these sources are not known, but it is
plausible that they are all  14.5~kpc away \citep{cwd+97, fmc+99}.  We believe that
even if \sgr\ is not directly associated with the LBV or \snr, it was
likely born in the same cluster, which requires that the progenitor of
\sgr\ was quite massive to have had a supernova before the LBV or the
stars in the cluster (as noted by \citealt{hkc+99}).  Perhaps this is
a case of an extremely high-mass star ($\gsim 50 M_{\sun}$) forming a neutron star
due to large mass loss over its lifetime \citep{fk01}?   Alternatively, the
sources could all have originated in the same molecular cloud, but the
supernova from \sgr\ could have triggered the star formation that lead
to the LBV and the cluster.  

The angular scale where the scattering halo departs from the psf is
smaller than that seen for \textit{ROSAT} data, and we do not measure
the halo out to such large radii, so the relations presented for
\textit{ROSAT} data by \citet{ps95} are not directly applicable.
However, we can 
correct the relation between fractional halo intensity and hydrogen
column density from \citet{ps95} to the appropriate energy and angular
scale range for our data, and  we find an expected halo intensity of
$\approx 0.3$\% based on the $N_{H}=\expnt{6}{22}\mbox{ cm}^{-2}$, as determined by
\citet{mcft00}.  This is very similar to the  measured intensity,
supporting the conclusion that 
we see a halo and not an instrumental artifact, and  implying that the size
distribution of the scattering grains at small angular scales is
similar to that seen for the larger \textit{ROSAT} halos.

The halo we see is very similar (in angular scale and fractional
intensity) to that seen for SGR~1900+14 \citep{ktw+01}, suggesting
that they lie behind similar dust columns (or possibly that both
``halos'' are in fact instrumental effects).  As with SGR~1900+14, the
pulsations of \sgr\ are too smeared out in the halo to use them to
determine a geometric distance \citep[cf.][]{ts73,pbpt00}.  However, if
observations were made of \sgr\ soon ($\sim$ hours to days) after a
significant change in the level of its emission (e.g.\ following a
giant flare) such that the change was visible in the halo, a distance
determination could be made. \textit{BeppoSAX} is probably the only
current satellite that has the necessary slew capabilities coupled
with spatial resolution and soft-energy response to perform these
observations.

\begin{figure*}
\plotone{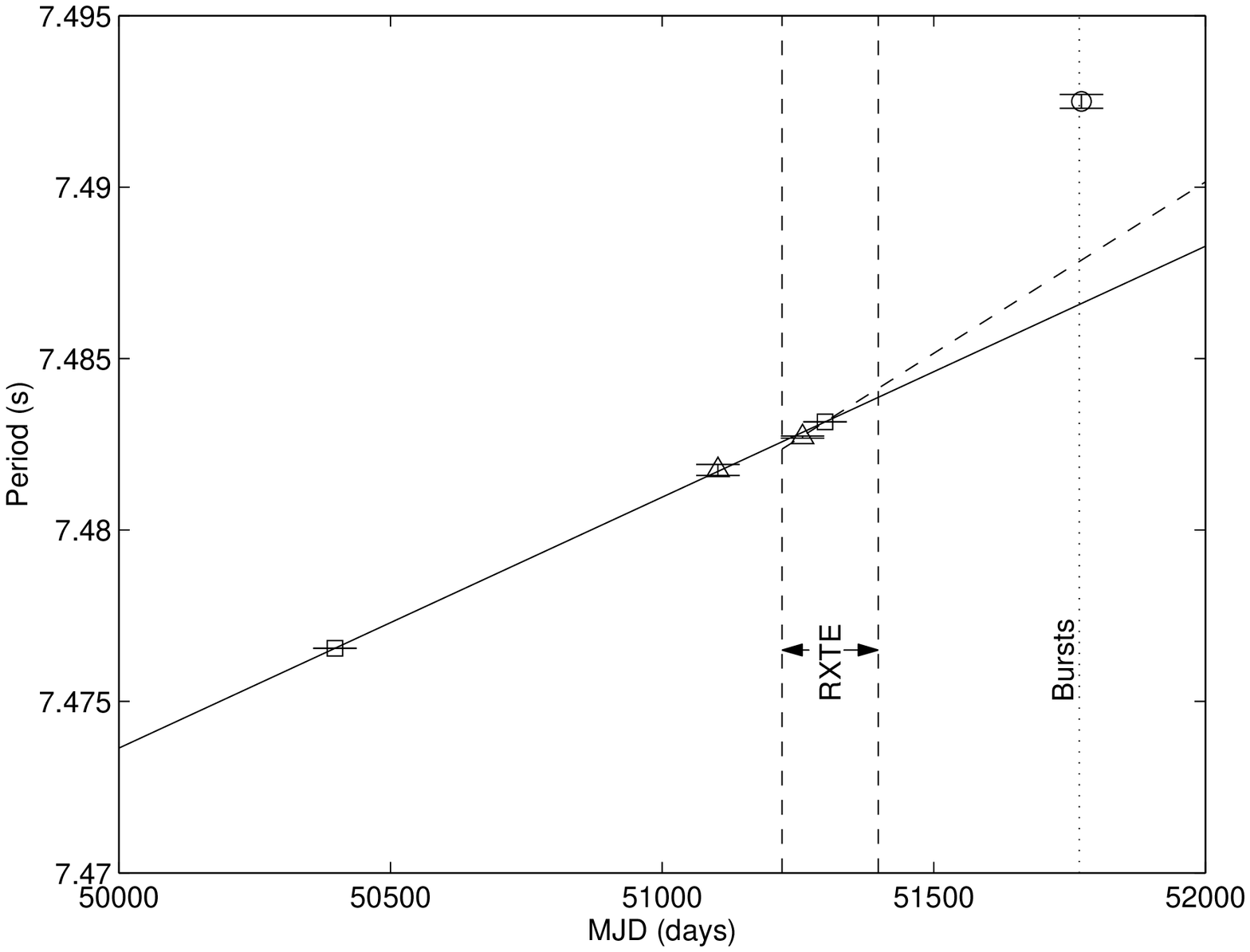}
\caption{Period evolution of \sgr, from \textit{RXTE} (squares;
\citealt{wkf+00}), \textit{BeppoSAX} (triangles; \citealt{mcft00}), and
\chandra\ (circle; this work).  The solid line is a linear
least-squares fit to the \textit{RXTE} and \textit{BeppoSAX} data with
$\dot{P}=\expnt{8.469(1)}{-11}\mbox{ s s}^{-1}$, from which the
\chandra\ data deviate significantly.  The vertical dashed lines
(labeled ``RXTE'') delimit the phase-connected \textit{RXTE}
ephemeris \citep{wkf+00}, which we then extrapolate as the diagonal
dashed line. 
The dotted vertical line marked ``Bursts'' indicates the time of the burst
activity that 
preceded the \chandra\ observation \citep{g+00}.
\label{fig:timing}}
\end{figure*}

The measured period of \sgr\ is not consistent with either its
long-term spin-down  \citep{wkf+00,mcft00} or the newer spin-down
measured from phase-connected \textit{RXTE} observations
\citep{wkf+00}, as seen in Figure~\ref{fig:timing}.
While \sgr\ exhibits substantial timing noise \citep{wkf+00}, the
long-term trend was generally stable  for a number of years.
However, \citet{wkf+00} found that at a single epoch the
instantaneous spin-down rate for \sgr\ was higher by a factor of
1.5 that the global trend, and to account for the new \chandra\  data,
the spin-down rate must have increased even more.   Between the last
\textit{RXTE} measurement from \citet{wkf+00} and our measurement, the
average spin-down rate is $\expnt{2.3(1)}{-10}\mbox{ s s}^{-1}$, which
is a factor of 2 higher than the largest $\dot{P}$ previously measured
for this source.  This change is similar to 
the large change in both spin period and spin-down rate seen for
SGR~1900+14 following its activation during 1998~August
(\citealt*{mrl99}; \citealt{wkvp+99}) and for the anomalous X-ray pulsar
1E~1048.1$-$5937 \citep{kgc+00}.    \sgr\ must have either 
increased its long-term spin-down substantially or suffered from a
momentary large change in spin-down.  Such sudden changes in spin-down are capable of being 
produced within the magnetar framework \citep{tdw+00}, either  due to
particle outflows or due to re-alignment of the neutron star crust,
but  they are typically assumed to be episodic and triggered by bursting or flaring
activity \citep{wkvp+99,tdw+00}.  We note that \citet{g+00}
did observe a period of moderate burst activity for \sgr\ several days
before the  31-ks \chandra\ observation, but this was too close to the
\chandra\ observation to have affected the spin-period significantly.
Whether sudden or gradual, the 
change in rotation for \sgr\ 
had to start before the bursts observed by
\citet{g+00} (unless the spin-down is 
orders of magnitude larger than previously seen), so these changes were not caused by burst
activity.  This suggests that the more gradual change is the correct
mechanism in this case, perhaps due to continuing plastic deformation
of the crust \citep[e.g.][]{tdw+00}.   In addition, spin-down cannot
be used to determine the ages or magnetic fields of any SGR in the manner
typical for radio pulsars \citep[e.g.][]{mt77}, as noted by
\citet{tdw+00}.  We must instead rely upon arguments relating to the
super-Eddington bursts \citep{td95}, global energetics, and similar
phenomena as indicators of the magnetar-like fields of these objects.

\section{Conclusions}
We have confirmed the position of \citet{hkc+99} for \sgr, and 
determined that it is near neither the non-thermal core of
\snr\ \citep{fvk97} nor the luminous blue variable star 
\citep[LBV;][]{vkkmn95}, although those two sources 
may be associated.  Through comparison of the radial profile of \sgr\
to a model psf, we see evidence for a broad X-ray scattering halo.
Given its violent nature, \sgr\ may produce changes in its X-ray
emission such that observations of this halo could lead to a geometric
distance determination, thereby fixing the luminosity scale of SGRs
and locating \sgr\ in three dimensions relative to the massive star
cluster \citep{fmc+99}, the LBV, and other  nearby objects.

Timing analysis of the \chandra\ data show that \sgr\ is not
continuing its long-term trend of spin-down that it has followed for
the past 4~years.  This is similar to changes in spin-down observed
for SGR~1900+14 following the 1998~August outbursts, although only
minor bursts were observed for \sgr\ around the time of these
observations and these bursts are unlikely to have been responsible
for the deviation from normal spin-down.

\acknowledgements We would like to thank R.~Rutledge and J.~Carpenter
for valuable conversations.  D.L.K.\ is supported by the Fannie and John Hertz
Foundation, S.R.K.\ by  NSF and NASA, and E.V.G.\ and G.V.\ by the NASA LTSA
grant NAG5-22250.  We have made
extensive use of the SIMBAD database, and would like to express our
appreciation of the astronomers who maintain this database.


\bibliographystyle{apj}



\begin{deluxetable}{l c c c c}
\tablecaption{Summary of \chandra\ Observations\label{tab_sum}}
\tablecolumns{4}
\tablewidth{0pt}
\tablehead{
\colhead{Date} & \colhead{MJD} & \colhead{Exposure} & \colhead{ACIS
S-3} & \colhead{TE Mode} \\
& \colhead{of Start} & \colhead{(ks)} & \colhead{Counts\tablenotemark{a}} & \\}
\startdata
2000~Jul~24 & 51749.7 & 4.9 & 833 & normal \\
2000~Aug~15 & 51771.8 & 31.1 & 7738 & $1/4$-subarray \\
\enddata
\tablenotetext{a}{Accepted counts (standard processing) within
$3\farcs3=10$-$\sigma$ radius.}  
\end{deluxetable}

\begin{deluxetable}{c c c c c c c}
\tablewidth{0pt}
\tablecaption{USNO Stars Detected in \chandra\ Images\label{tab:usno}}
\tablehead{
\colhead{Star} & \mc{3}{c}{X-ray Position} & 
\mc{2}{c}{USNO Position\tablenotemark{a}} & \colhead{$\Delta r$} \\
 & \colhead{$\alpha_{2000}$} & \colhead{$\delta_{2000}$}  &
\colhead{$\sigma$\tablenotemark{b}} &
\colhead{$\alpha_{2000}$} & \colhead{$\delta_{2000}$} & \\
 & & & \colhead{(arcsec)} & & & \colhead{(arcsec)} \\
}
\startdata
\mc{7}{c}{Sources in 5-ks Image} \\
A & $18^{\rm h}08^{\rm m}24\fs93$  & $-20\degr24\arcmin33\farcs2$ &
0.2 &
$18^{\rm h}08^{\rm m}24\fs95$ & $-20\degr24\arcmin32\farcs6$ &  0.7 \\
B & $18^{\rm h}08^{\rm m}43\fs41$  & $-20\degr23\arcmin58\farcs8$  &
0.3 &
$18^{\rm h}08^{\rm m}43\fs46$ & $-20\degr23\arcmin58\farcs1$ &  1.0 \\
\mc{7}{c}{Sources in 31-ks Image} \\
B & $18^{\rm h}08^{\rm m}43\fs43$ & $-20\degr23\arcmin58\farcs6$ &
 0.1 & 
 $18^{\rm h}08^{\rm m}43\fs46$ & $-20\degr23\arcmin58\farcs1$ & 0.6 \\
C & $18^{\rm h}08^{\rm m}41\fs43$ & $-20\degr25\arcmin12\farcs8$ &
 0.3 & 
 $18^{\rm h}08^{\rm m}41\fs47$ &  $-20\degr25\arcmin12\farcs6$ &  0.6 \\
\enddata
\tablenotetext{a}{Uncertainties on USNO positions are assumed to be
$0\farcs2$ rms in each coordinate.}
\tablenotetext{b}{RMS uncertainty in each coordinate.}
\end{deluxetable}


\end{document}